\documentstyle[12pt,epsfig]{article}
\voffset = - 0.10 in 
\begin{document}
\thispagestyle{empty}

%\hfill {\large\bf DRAFT}

%\hfill {\large\bf \today}

%\vspace*{1cm}

\begin{center}
{\Large\bf 
Charged Pion Energy Reconstruction in the ATLAS Barrel
Calorimeter 
}
\end{center}

\bigskip

\begin{center}
{\large\bf M.\ Bosman}

\smallskip

{\it 
Universitat Aut\`onoma de Barcelona\\
Institut de F\'{\i}sica d'Altes Energies\\
Barcelona, Spain
}

\bigskip

{\large\bf Y.\ Kulchitsky}

\smallskip

{\it
JINR, Dubna, Russia\\
\&
Institute of Physics,  \\
National Academy of Sciences, Minsk, Belarus
}

\bigskip

{\large\bf M.\ Nessi} 

\smallskip

{\it
CERN, Geneva, Switzerland
}
\end{center}

%\vspace*{\fill}

\begin{abstract}
The intrinsic performance of the ATLAS barrel and extended barrel 
calorimeters for the measurement of charged pions is presented. 
Pion energy scans ($E = 20,\ 50,\ 200,\ 400$ and $1000$ GeV) 
at two pseudo-rapidity points ($\eta = 0.3$ and $1.3$) and
pseudorapidity scans ($-0.2 < \eta < 1.8$) with pions of constant
transverse energy ($E_T = 20$ and $50$ GeV) are analysed. 
A simple approach, that accounts in first order for non-compensation and
 dead material effects, is used for the pion energy reconstruction. 
The intrinsic performances of the calorimeter are studied: resolution,
 linearity, effect of dead material,  tails in the energy distribution.
The effect of electronic noise, cell energy cuts and restricted cone size
are investigated.
\end{abstract}

\newpage

%%%%%%%%%%%%%%%%%%%%%%%%%%%%%%%%%%%%%%%%%
\section{ATLAS Calorimetry}
\begin{figure}[tbph]
\begin{center}
\mbox{\epsfig{figure=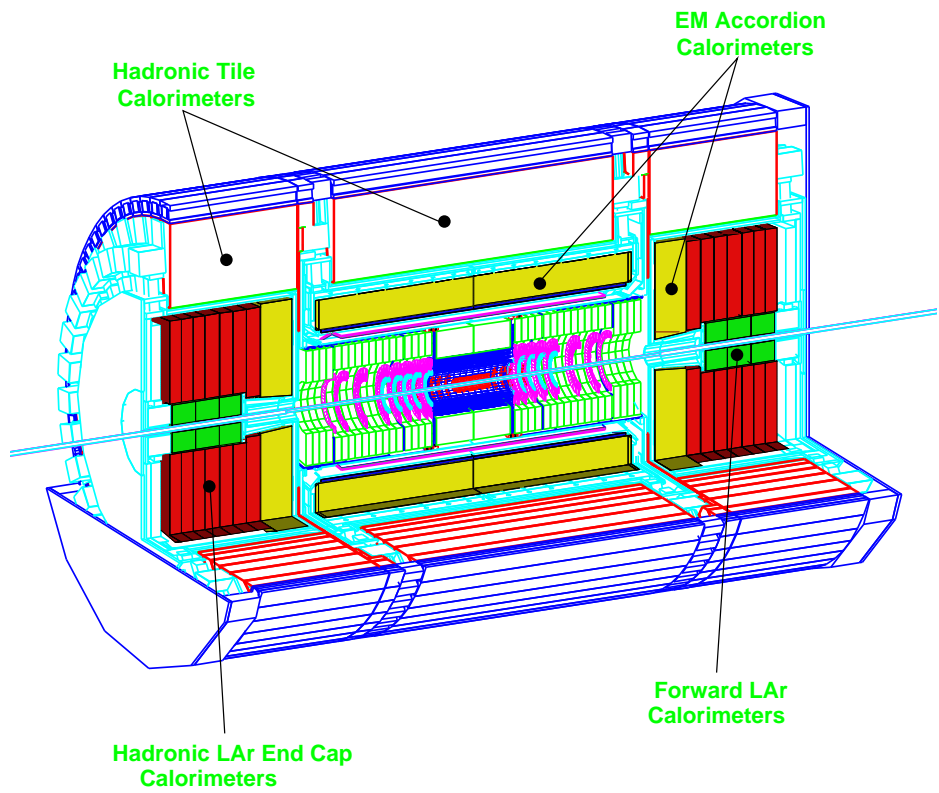,width=1.0\textwidth}} 
 \end{center}
 \caption{ 
        Three-dimensional cutaway view of the ATLAS calorimeters.
        }
\label{f-00}
\end{figure}

%%%%%%%%%%%%%%%%%%%%%%%%%%%%%%%%%%%%%%%%%
A view of the ATLAS calorimeters \cite{tdr1} is presented in 
Figure \ref{f-00}.
The calorimetry consists of an electromagnetic (EM) calorimeter covering 
the pseudorapidity region $|\eta| < 3.2$, a hadronic barrel calorimeter
covering $|\eta| < 1.7$, hadronic end-cap calorimeters covering
$1.5 < |\eta| < 3.2$,  and forward calorimeters covering 
$3.1 < |\eta| < 4.9$.

The EM calorimeter is a lead/liquid-argon (LAr) detector with accordion 
geometry \cite{tdr2}.
Over the pseudorapidity range $|\eta| < 1.8$, it is preceded by 
a presampler detector, installed immediately behind the cryostat cold wall,
and used to correct for the energy lost in the material upstream of the 
calorimeter.

The hadronic calorimetry of ATLAS, presented in Figure \ref{f-00}, 
consists of three main devices. 
In the barrel region ($ |\eta| < 1.7$) 
there is the scintillating Tile Calorimeter \cite{tdr3}. 
The Hadronic End-cap LAr Calorimeter (HEC) extends up to $|\eta| = 3.2$. 
The range $3.1 < |\eta| < 4.9$
is covered by the high density Forward Calorimeter (FCAL). 
Up to $|\eta|= 2.5$ the basic granularity of the hadron calorimeters is 
$\Delta \eta \times \Delta \phi = 0.1 \times 0.1$. 
This region is used for precise measurements of the energy and angles 
of jets. 
In the region $|\eta| > 2.5$, the basic granularity is approximately 
$\Delta \eta \times \Delta \phi = 0.2 \times 0.2$.
 
A more detailed description of all ATLAS calorimeters is given in the 
Calorimeter TDRs (\cite{tdr1}, \cite{tdr2} and \cite{tdr3}).

The performance of the barrel and extended barrel
sections of the ATLAS hadronic calorimeter for the measurement of charged 
pion energy is studied. 
The intrinsic energy resolution, the effects of dead material,
 electronic noise and limited cone size are discussed. 

%%%%%%%%%%%%%%%%%%%%%%%%%%%
\section{Energy Resolution}
%%%%%%%%%%%%%%%%%%%%%%%%%%%
In the barrel region, the response of the calorimeter was studied at 
two pseudorapidity values: 
$\eta = 0.3$ (central barrel) and 
$\eta = 1.3$ (extended barrel). 
First, the energy sampled in the different calorimeter compartments is 
converted to the total deposited energy using the electromagnetic energy 
scale (EM scale). 
The intrinsic performance of the calorimeter is studied: 
the energy considered is not restricted to a cone and electronic 
noise is not added. 
These effects are discussed later in Section \ref{noise}. 
The algorithm to reconstruct the pion energy is similar to the
 ''Benchmark Method''  used to analyse the combined LAr-Tile test beam data 
\cite{tilecal-nima387}
%(see Section 5.1.1 and Equation 5-1):
\begin{equation}
\label{eq-03}
        E_{rec} = 
          \alpha \cdot E_{had} 
        + \beta \cdot E_{em} 
        + \gamma E_{em}^2
        + \delta \cdot \sqrt{E_{had\ 1} \cdot E_{em\ 3}}
        + \kappa \cdot E_{ITC}
        + \lambda \cdot E_{scint}\ . 
\end{equation}
        
\begin{figure}[tbph]
\begin{center}
\mbox{\epsfig{figure=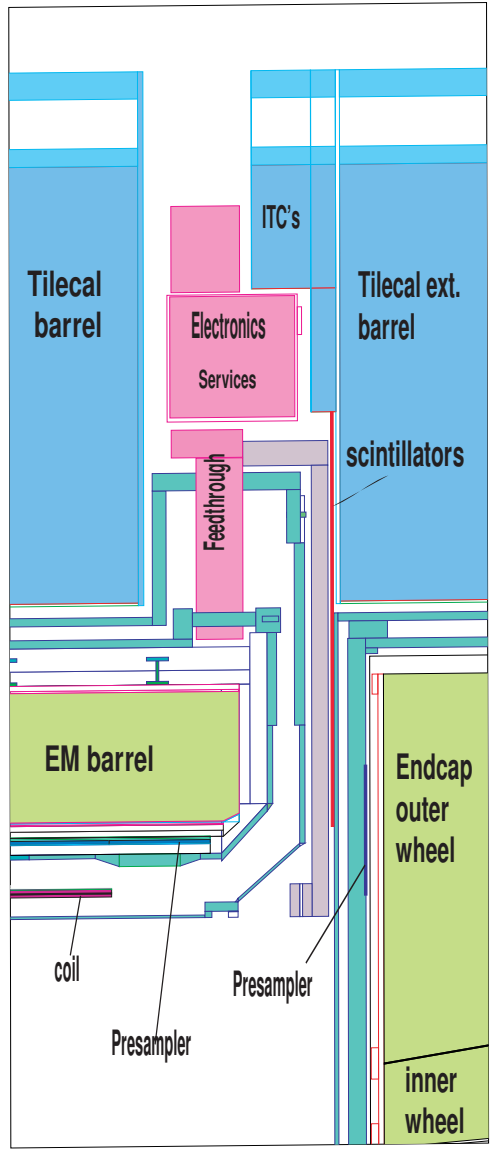,width=0.9\textwidth,height=0.9\textheight}} 
\end{center}
 \caption{ 
        Longitudinal view of a quadrant of the vertical gap between the 
        barrel and extended barrel Hadronic Tile Calorimeter.
        }
\label{f-01}
\end{figure}

The coefficients $\alpha$ and $\beta$ take into account the different
response of the 
EM and Hadronic Calorimeters to the pion energy. 
The quadratic term $\gamma \ E_{em}^2$ 
provides an additional first order correction for non-compensation
(the coefficient is negative, it suppresses the signal for events with a
large fraction of electromagnetic energy). 
The term $\delta \ \sqrt{E_{had\ 1} \cdot E_{em\ 3}}$
estimates the energy loss in the cryostat wall separating the
LAr and Tile Calorimeters. 
In the central barrel, the energy is taken from the geometric mean of the 
energies in the last compartment of the LAr EM barrel
($E_{em\ 3}$) and the first 
compartment of the Tile barrel calorimeter ($E_{had\ 1})$;
 whereas in the extended barrel 
the energy is taken from the geometric mean of the energies in the outer 
wheel of the EM end-cap and the first compartment of the Tile extended 
barrel calorimeter. 
The term $\kappa \ E_{ITC}$
corrects for the energy loss in the dead material in the vertical 
gap between the Tile central and extended barrels. 
It is sampled by the two 
Intermediate Tile Calorimeter (ITC) modules (see Figure 
% Martine, please give me ``eps'' file for FIG. 5-i from Physics TDR 
\ref{f-01}). 
The last term $\lambda \ E_{scint}$
corrects for the energy loss in the barrel and end-cap 
vertical cryostat walls (see Figure \ref{f-01}).

\begin{figure}[tbph]
\begin{center}
\begin{tabular}{c}
\mbox{\epsfig{figure=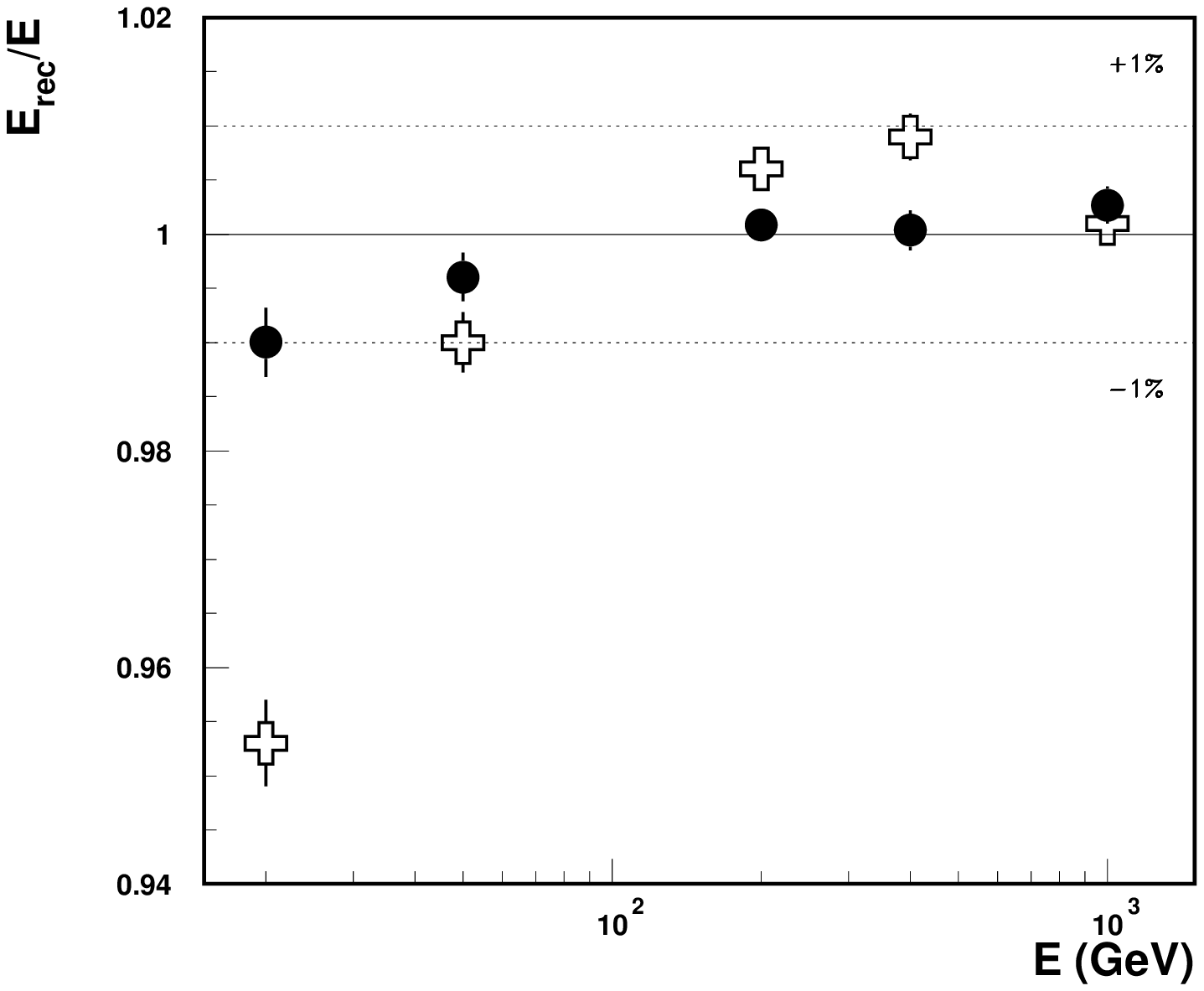,width=0.8\textwidth,height=0.4\textheight}}
\\
\mbox{\epsfig{figure=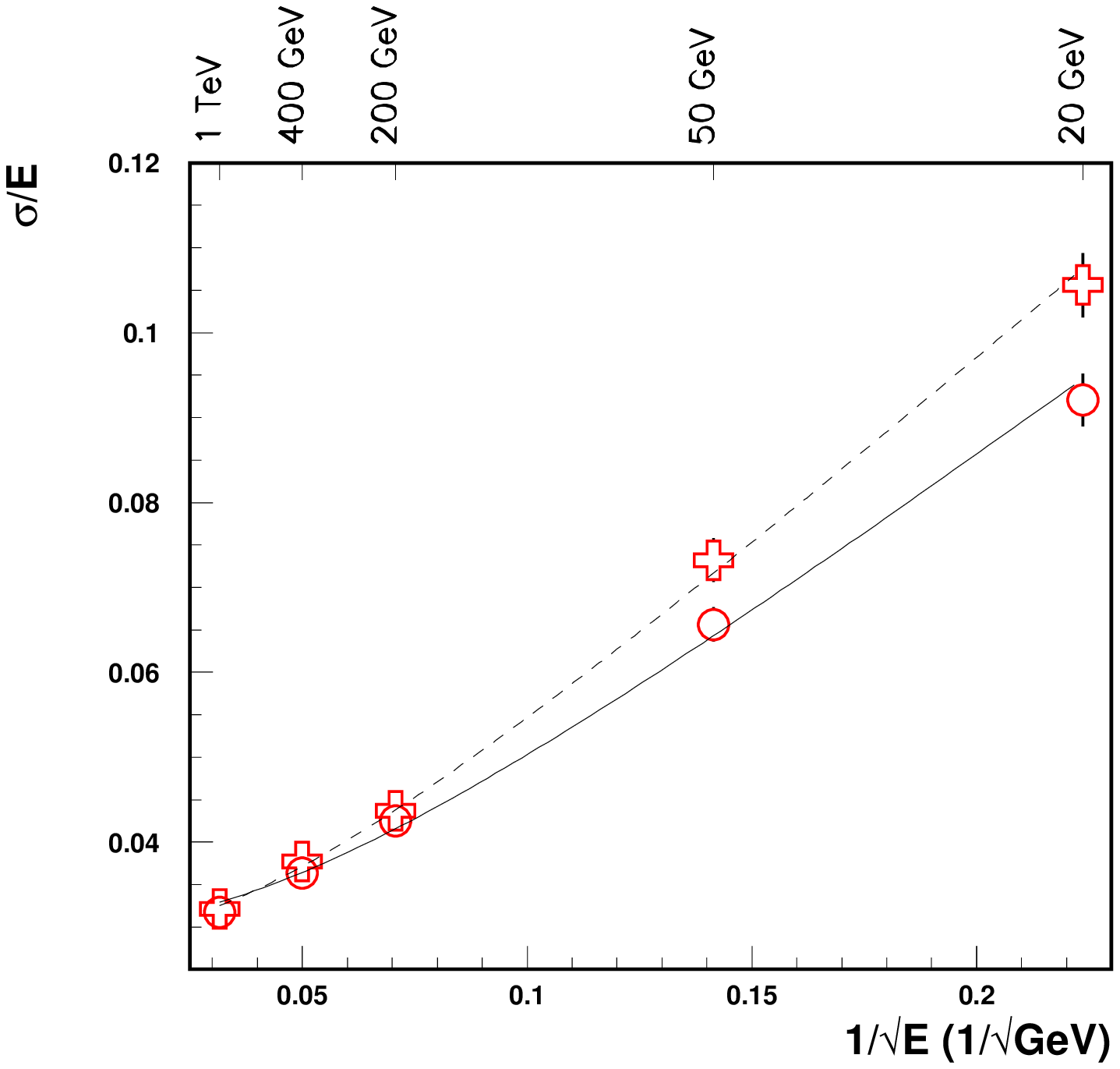,width=0.8\textwidth,height=0.4\textheight}}
\\
\end{tabular}
\end{center}
 \caption{
        Pion energy scan in the central barrel ($\eta = 0.3$).
        The top plot shows the residual non-linearity,
        the bottom plot shows the energy resolution with the results 
        of the fit with Equation $\sigma / E = a/\sqrt{E} \oplus b$.
        Two sets of parameters for the pion energy reconstruction 
        have been used: open crosses --- for energy independent parameters;
        solid dots --- for parameters fitted at each energy and 
        pseudorapidity.  
        }
\label{f-02}
\end{figure}

\begin{figure}[tbph]
\begin{center}
\begin{tabular}{c}
\mbox{\epsfig{figure=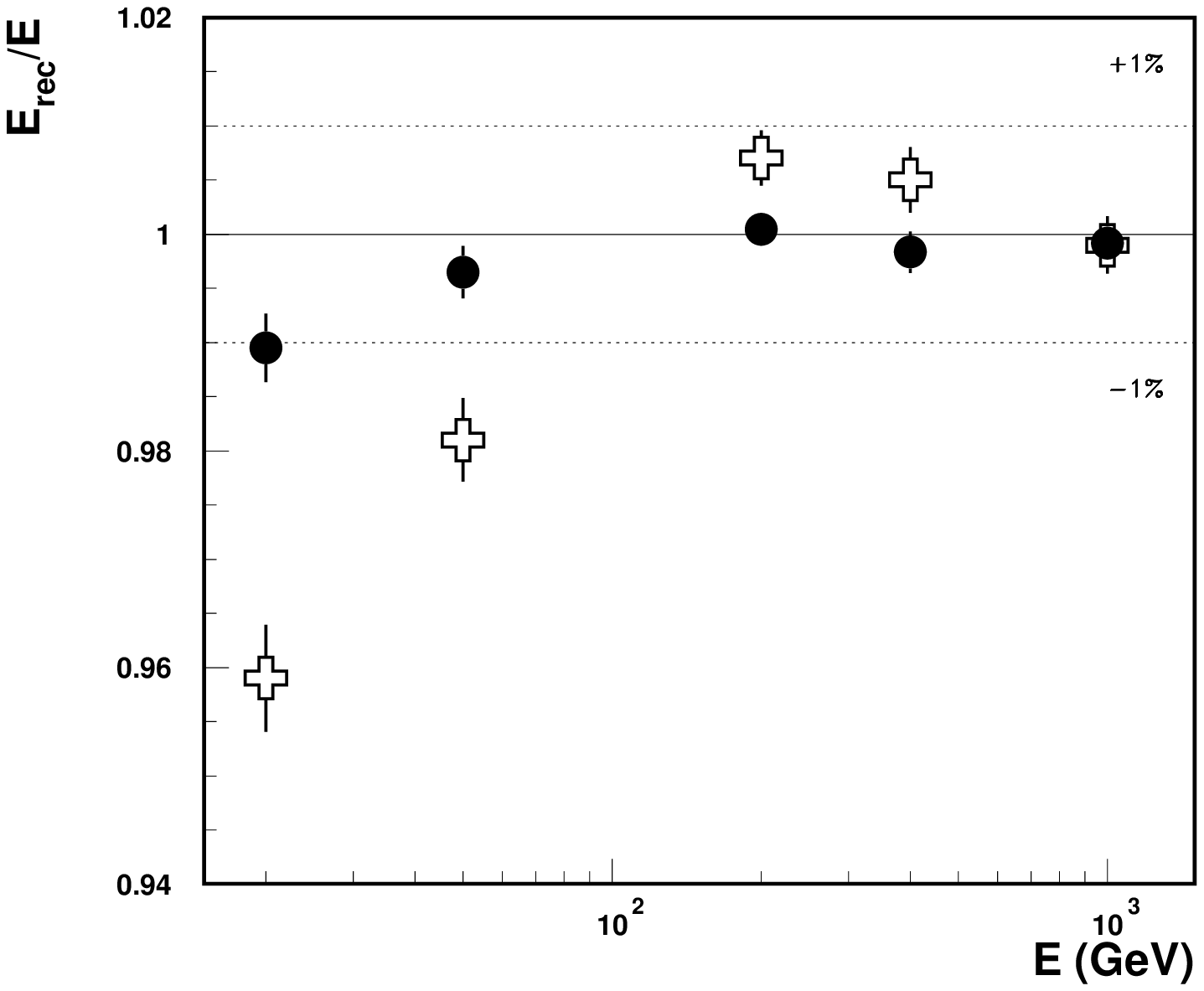,width=0.8\textwidth,height=0.4\textheight}}
\\
\mbox{\epsfig{figure=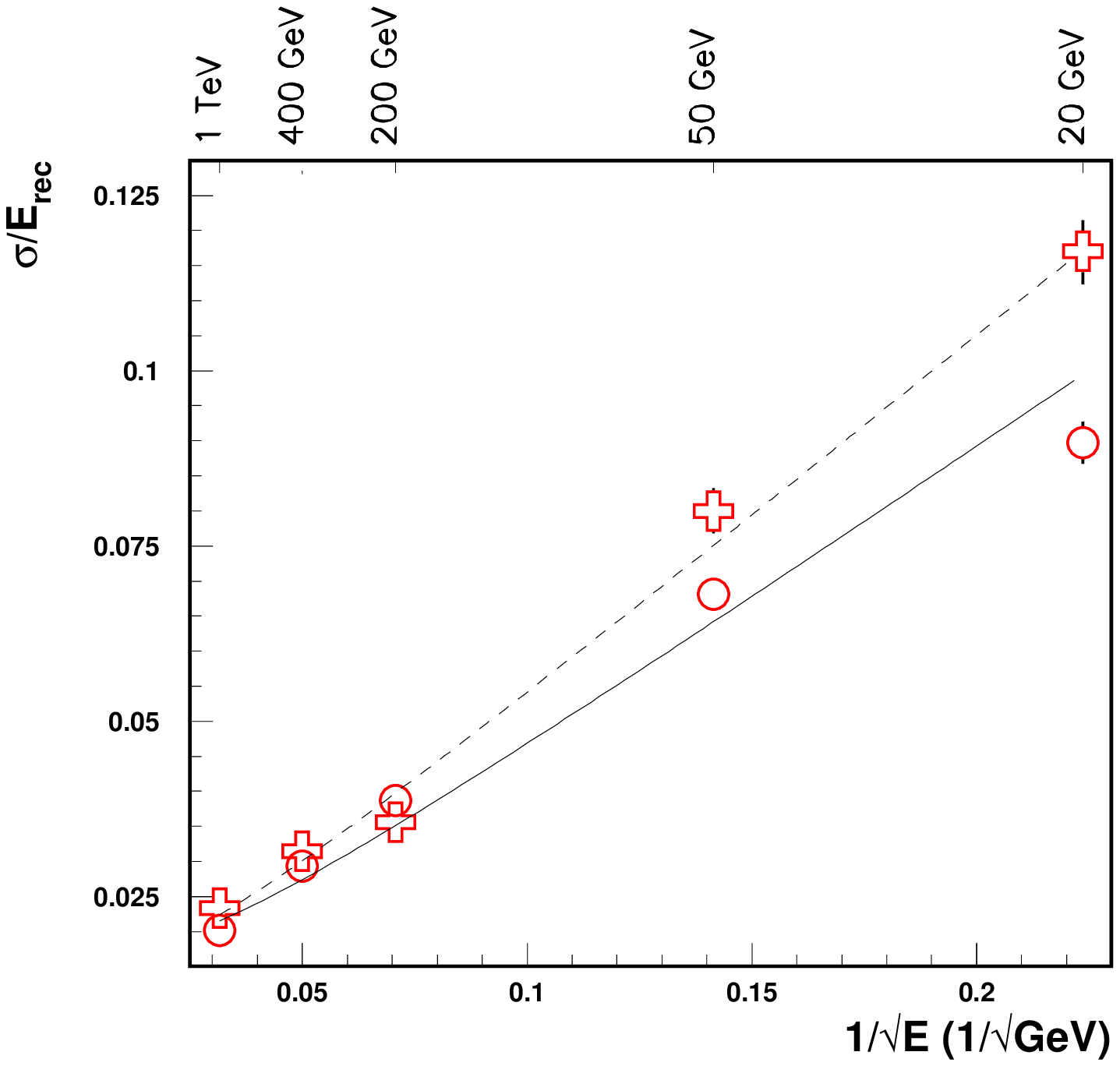,width=0.8\textwidth,height=0.4\textheight}}
\\
\end{tabular}
\end{center}
 \caption{ 
        Pion energy scan in the extended barrel ($\eta = 1.3$).
        The top plot shows the residual non-linearity,
        the bottom plot shows the energy resolution with the results 
        of the fit with Equation $\sigma / E = a/\sqrt{E} \oplus b$.
        Two sets of parameters for the pion energy reconstruction 
        have been used: open crosses --- for energy independent parameters;
        solid dots --- for parameters fitted at each energy and 
        pseudorapidity.         
        }
\label{f-03}
\end{figure}

The response and the energy resolution for pions in the energy range from 
$E_0 = 20\ GeV$ to $1\ TeV$ at $\eta = 0.3$ and 1.3 are shown in 
Figures \ref{f-02} and \ref{f-03}.
The open crosses show the results when the coefficients of Equation 
(\ref{eq-03}) are independent of energy.
With the simple ''Benchmark Method'', the effect of non-compensation
is not fully corrected for and
there is a residual non-linearity of the pion response of the order of
$4 - 5 \%$ between 20 GeV and 1 TeV. The test beam data show a
10\% residual non-linearity between 20 and 300 GeV when using the same 
reconstruction method \cite{tilecal-nima387},
reflecting the fact that G-CALOR \cite{G-CALOR} 
predicts a lower degree of non-compensation and
may not describe correctly the energy dependence of the fraction of 
electromagnetic energy produced in the pion interaction.
The energy dependence of the resolution is fitted with the two-term formula
\begin{equation}
\label{eq-04}
        \sigma / E = a/\sqrt{E} \oplus b \ 
\end{equation}
where the sampling term $a$ is given in $\%\sqrt{GeV}$ and the constant 
term $b$ in $\%$. 
Although the resolutions obtained for low-energy pions are similar 
in both cases, at high energy there is some longitudinal leakage 
in the central barrel, yielding a resolution at 1 TeV of 3\%
instead of the 2\% obtained in the extended barrel.
When energy dependent parameters are applied (solid dots),
the linearity of the response is restored
\footnote{The 1\% residual non-linearity at 20 GeV results from the fact
that the coefficients were obtained by minimizing the expression
$\sum{(E_{rec}-E_0})^2$ without the addition of a linear term
$\sum{(E_{rec}-E_0})$ with a Lagrange multiplier.} 
and the resolution improved.
The results are presented in  Table \ref{tb-01}:
%1
\begin{table}[tbph]
\begin{center}
\caption{
        Terms of the pion energy resolution 
        for the $\eta = 0.3$ and $1.3$
        (results of the fit with Equation (\ref{eq-04})).
        }
\label{tb-01}
\begin{tabular}{|c|c|c|}
\multicolumn{3}{l}{\mbox{~~~}}          \\
\hline
$\eta$  &$a$ ($\% \ \sqrt{GeV}$)&$b$ ($\%$)     \\ 
\hline
0.3     & $40\pm1$              &$3.0\pm0.1$    \\ 
\hline
1.3     & $44\pm3$              &$1.6\pm0.3$    \\ 
\hline
\end{tabular}
\end{center}
\end{table}

%%%%%%%%%%%%%%%%%%%%%%%%%%%%%
\section{Pseudorapidity Scan}
%%%%%%%%%%%%%%%%%%%%%%%%%%%%%
A pseudorapidity scan with pions of constant transverse energy 
$E_T = 20$ and 50 GeV was carried out to check that the linearity of 
the response can be maintained across the pseudorapidity range covered by
the barrel and the extended barrel, 
and that no significant tail appears 
in the line shape. 
The algorithm, characterised by 
Equation (\ref{eq-03}), 
with energy dependent parameters was applied. The parameters were adjusted
independently for the six sets of pion data, each one covering an 
interval of 0.4 in pseudorapidity.

\begin{figure}[tbph]
\begin{center}
\begin{tabular}{c} 
\mbox{\epsfig{figure=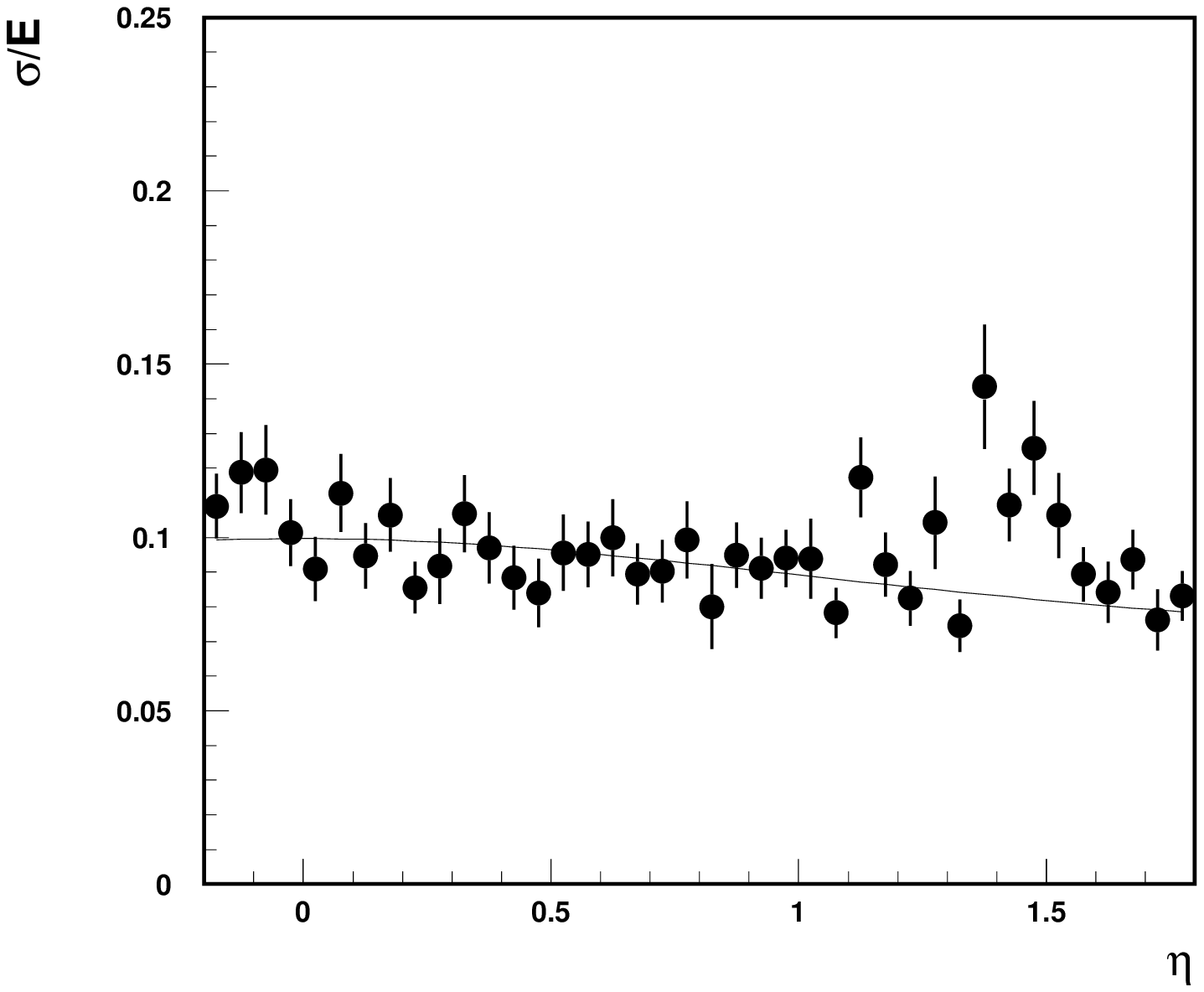,width=0.8\textwidth,height=0.4\textheight}}
\\
\mbox{\epsfig{figure=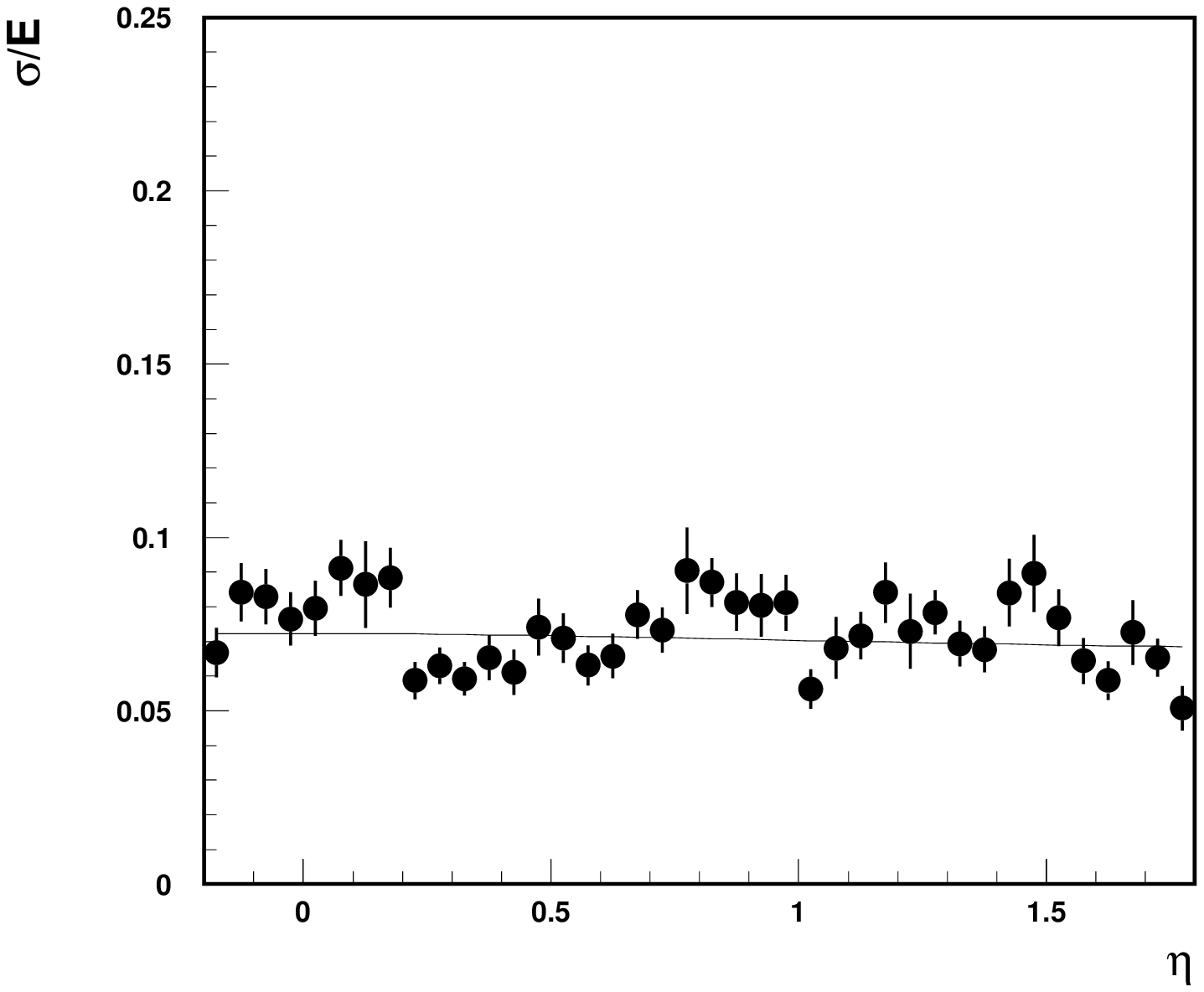,width=0.8\textwidth,height=0.4\textheight}}
\\
\end{tabular}
\end{center}
 \caption{ 
        The dependence of the energy resolution on pseudorapidity 
        for charged pions of constant transverse energy:
        $E_T = 20$ GeV (top plot) and $E_T = 50$ GeV (bottom plot).
        The lines correspond to the energy resolution parametrised using
        Equation $\sigma / E = 39\%/\sqrt{E} \oplus 1\%$ for $E_T = 20$ GeV 
        and $\sigma / E = 49\%/\sqrt{E} \oplus 2\%$ for $E_T = 50$ GeV.
        }
\label{f-04}
\end{figure}

The energy resolutions obtained for the two scans are shown in Figure 
\ref{f-04}. 
The solid lines show the energy resolution corresponding to 
Equation $\sigma / E = 39\%/\sqrt{E} \oplus 1\%$ for $E_T = 20$ GeV 
and $\sigma / E = 49\%/\sqrt{E} \oplus 2\%$ for $E_T = 50$ GeV.
This performance allows to fulfil the goal 
 for the jet energy resolution of the ATLAS 
hadronic calorimetry in the region $|\eta| < 3$ of 
$\sigma / E = 50\%/\sqrt{E} \oplus 3\%$. 
 
In the region of the cracks between the calorimeters,
from about $|\eta|= 1.3$ to $|\eta| = 1.5$, 
where the amount of dead material is the largest, the resolution is somewhat 
worse. 

\begin{figure}[tbph]
\begin{center}
\begin{tabular}{c}
\mbox{\epsfig{figure=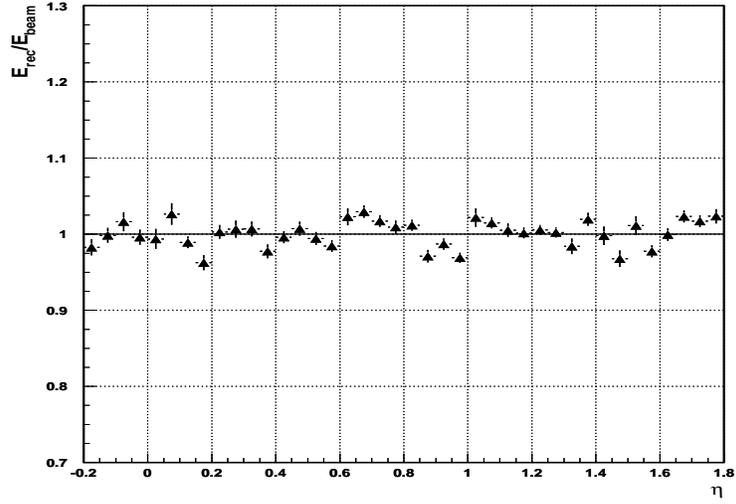,width=0.8\textwidth,height=0.4\textheight}}
\\
\mbox{\epsfig{figure=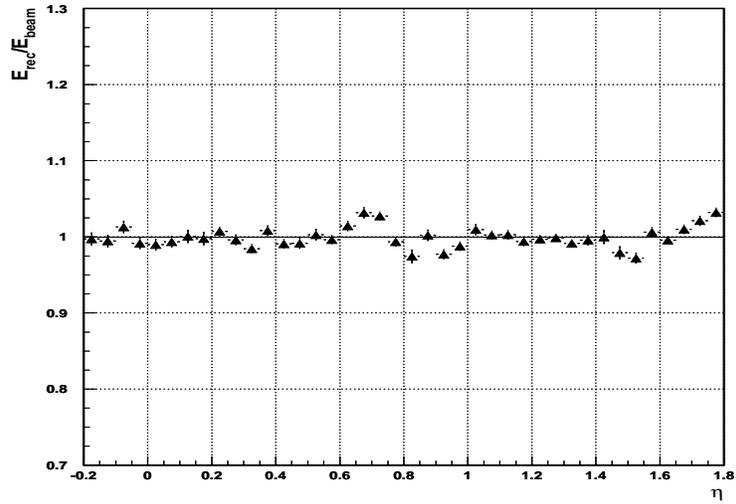,width=0.8\textwidth,height=0.4\textheight}}
\\
\end{tabular}
\end{center}
 \caption{ 
        Single charged pion linearity across $\eta$:
        $E_T = 20$ GeV (top plot) and $E_T = 50$ GeV (bottom plot).
        Calibration coefficients are adjusted independently for each 0.4 bin
        in $\eta$ but are kept constant within that interval. The figure
        shows the fitted mean of the pion distribution per bin of 0.05 in
        $\eta$.
        }
\label{f-04b}
\end{figure}

Figure \ref{f-04b} shows the linearity of the response across $\eta$.
The fitted mean is plotted for each interval of 0.05 in $\eta$. The RMS of
 the mean is 1.1\% for  $E_T = 50$ GeV and 2.0\% for $E_T = 20$ GeV.

\begin{figure}[tbph]
\begin{center}
\begin{tabular}{c}
\mbox{\epsfig{figure=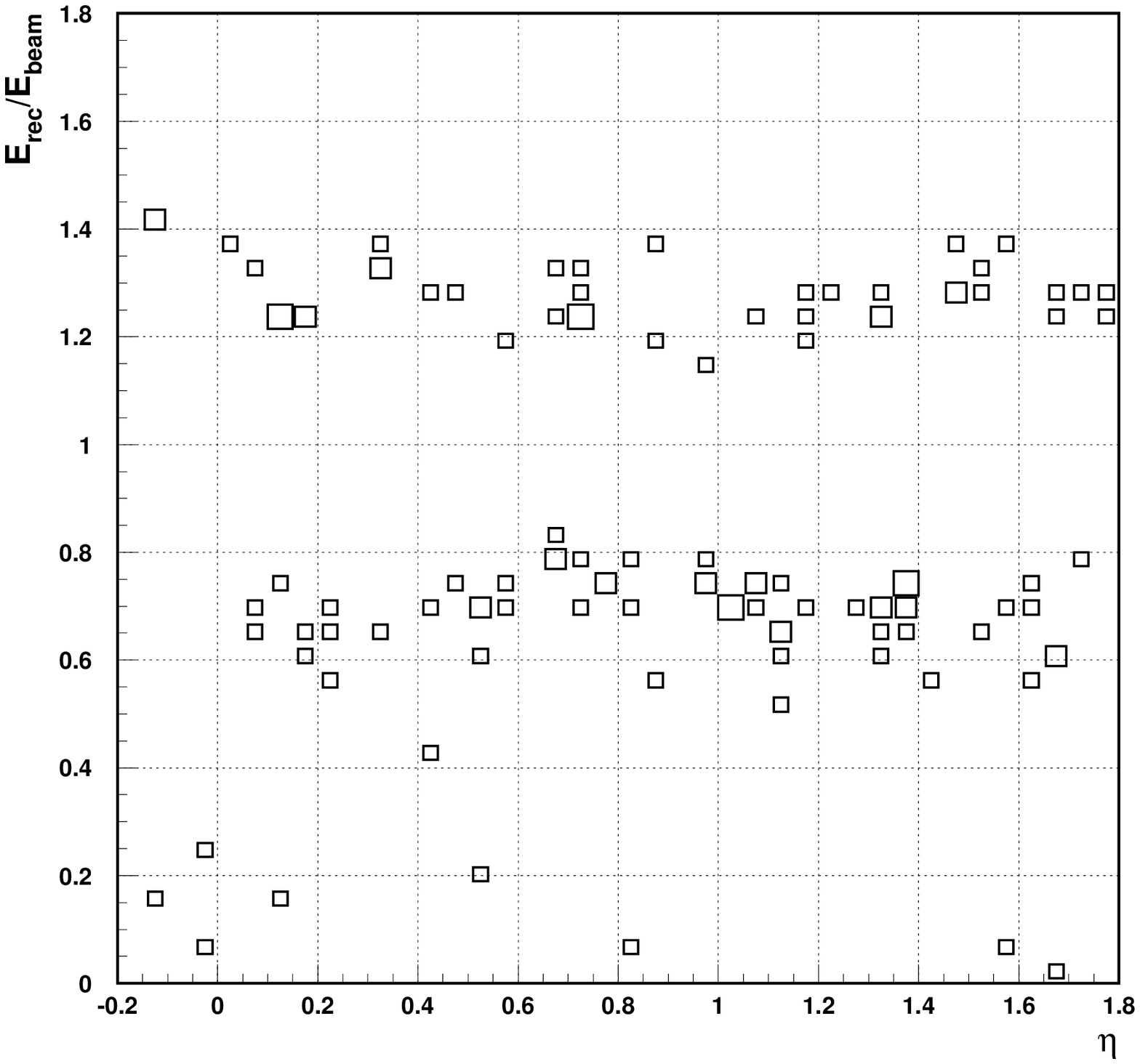,width=0.8\textwidth,height=0.4\textheight}}
\\
\mbox{\epsfig{figure=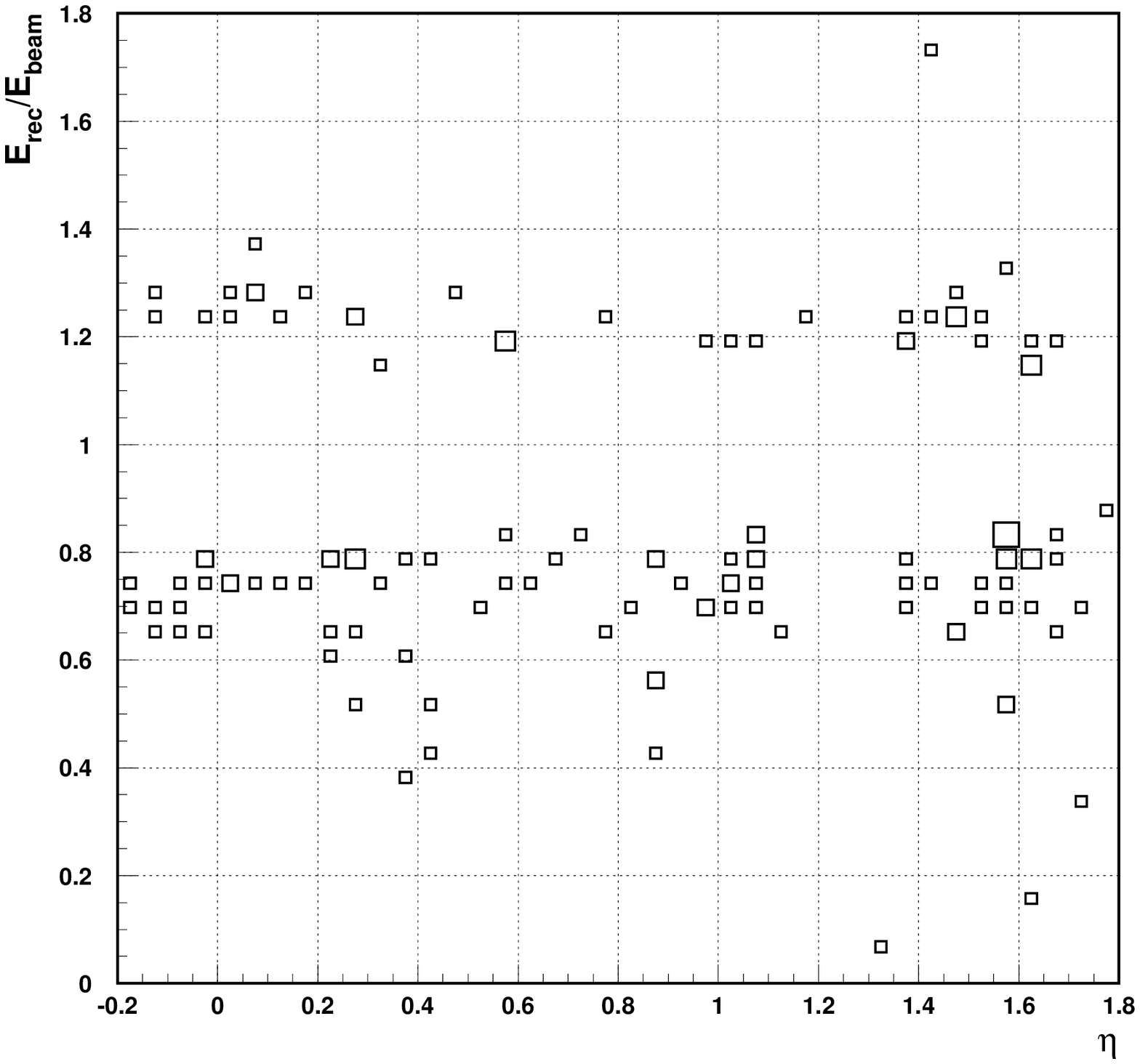,width=0.8\textwidth,height=0.4\textheight}}
\\
\end{tabular}
\end{center}
 \caption{
        Events in the tails of the distribution of the reconstructed 
        energy as a function of pseudorapidity for pions of  
        $E_T = 20$ GeV (top plot) and $E_T = 50$ GeV (bottom plot).
        Tails are defined as events with reconstructed energies more than
        three standard deviations away from the mean. 
        }
\label{f-05}
\end{figure}

In addition, the tails of the distributions of the reconstructed energy 
were investigated. 
Figure \ref{f-05} 
shows the events with a pion response more than three standard deviations 
away from the mean. 
No significant tails are present: 
the fraction of events in the tails does not exceed  1 -- 2\%. 
A few events out of a total of 5000 events per energy scan, mostly from 
the sample of pions of $E_T = 20$ GeV, deposit relatively little energy. 
These correspond to pions decaying to muons before reaching the calorimeter.  

%%%%%%%%%%%%%%%%%%%%%%%%%%%%%%%%%%%%%%%%%%%%%%%%%%%
\section{Effects of Electronic Noise and Cone Size}
\label{noise}
%%%%%%%%%%%%%%%%%%%%%%%%%%%%%%%%%%%%%%%%%%%%%%%%%%%
The results presented so far were obtained without any restriction on the 
pion reconstruction volume. 
These results characterise the intrinsic performance of the calorimeters. 
The presence of electronic noise does not allow integration over a too wide  
region, therefore the measurement of the pion energy must be restricted to 
a cone 
\begin{equation}
\label{eq-05}
        \Delta R = \sqrt{\Delta^2 \eta + \Delta^2 \phi} \ .
\end{equation}
 
A compromise has to be found between the pion energy lost outside of this 
cone and the noise included inside. 
The optimum varies as a function of pseudorapidity, since the showers have 
a width which is characterised by the polar angle whereas the calorimeter 
cells subtend intervals of constant pseudorapidity. 
Hence, at higher values of pseudorapidity, the showers extend laterally 
over more cells. 

For a cone of $\Delta R = 0.6$ ($\Delta R = 0.3$), 
the noise is above 3 GeV (1.5 GeV). 
Digital filtering \cite{cleland-nima338}
allows noise suppression (approximately by a factor 1.6). 
But even this level of noise is large and is comparable to the intrinsic 
resolution of the calorimeters for pions with energy of a few tens of GeV. 
A smaller cone of $\Delta R = 0.3$ 
is preferable from this point of view; 
after digital filtering, noise can be kept around 1 GeV in the barrel region 
and below 3 GeV in the pseudorapidity region covered by the extended barrel. 

\begin{figure}[tbph]
\begin{center}
\begin{tabular}{c}
\mbox{\epsfig{figure=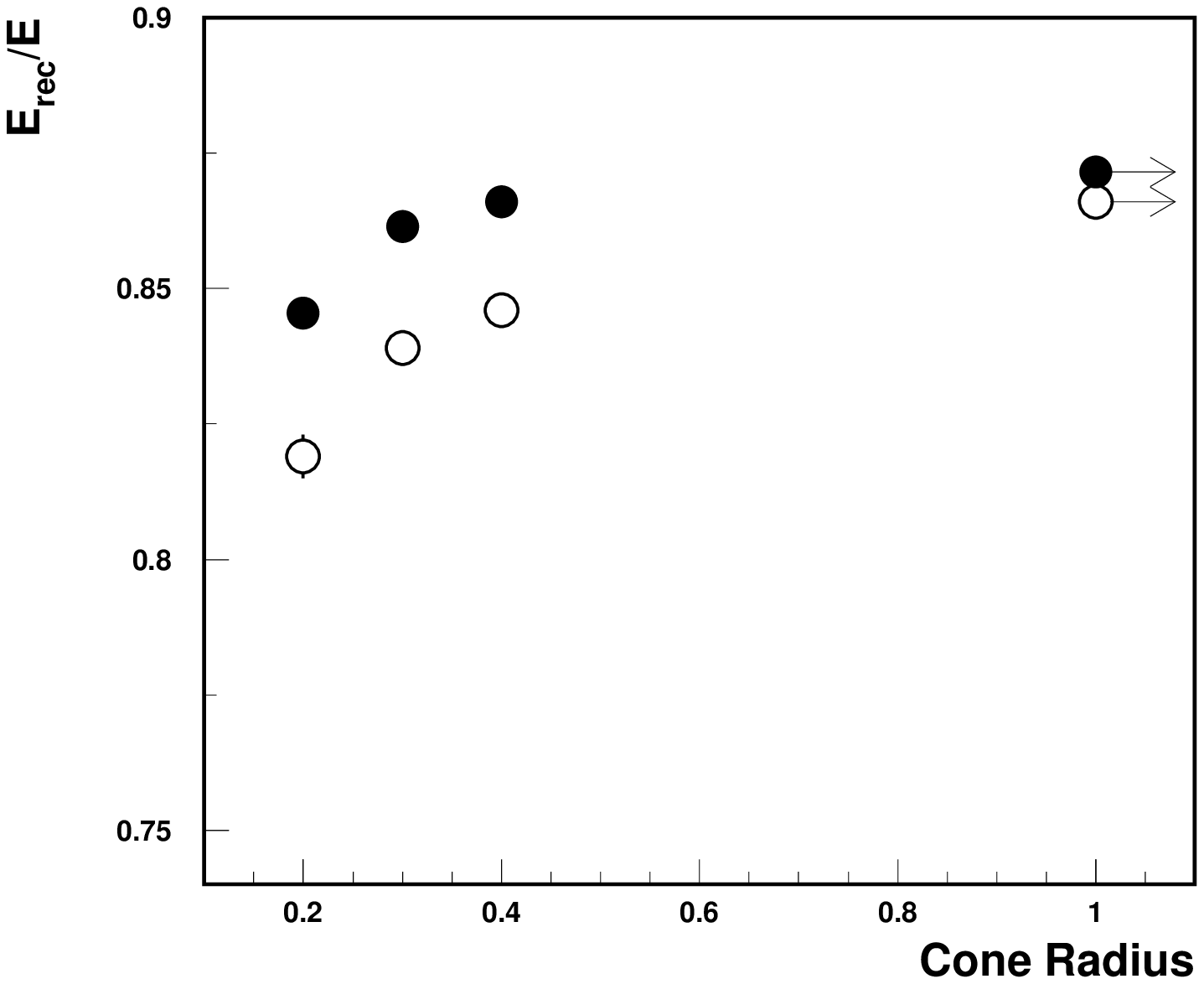,width=0.8\textwidth,height=0.4\textheight}}
\\
\mbox{\epsfig{figure=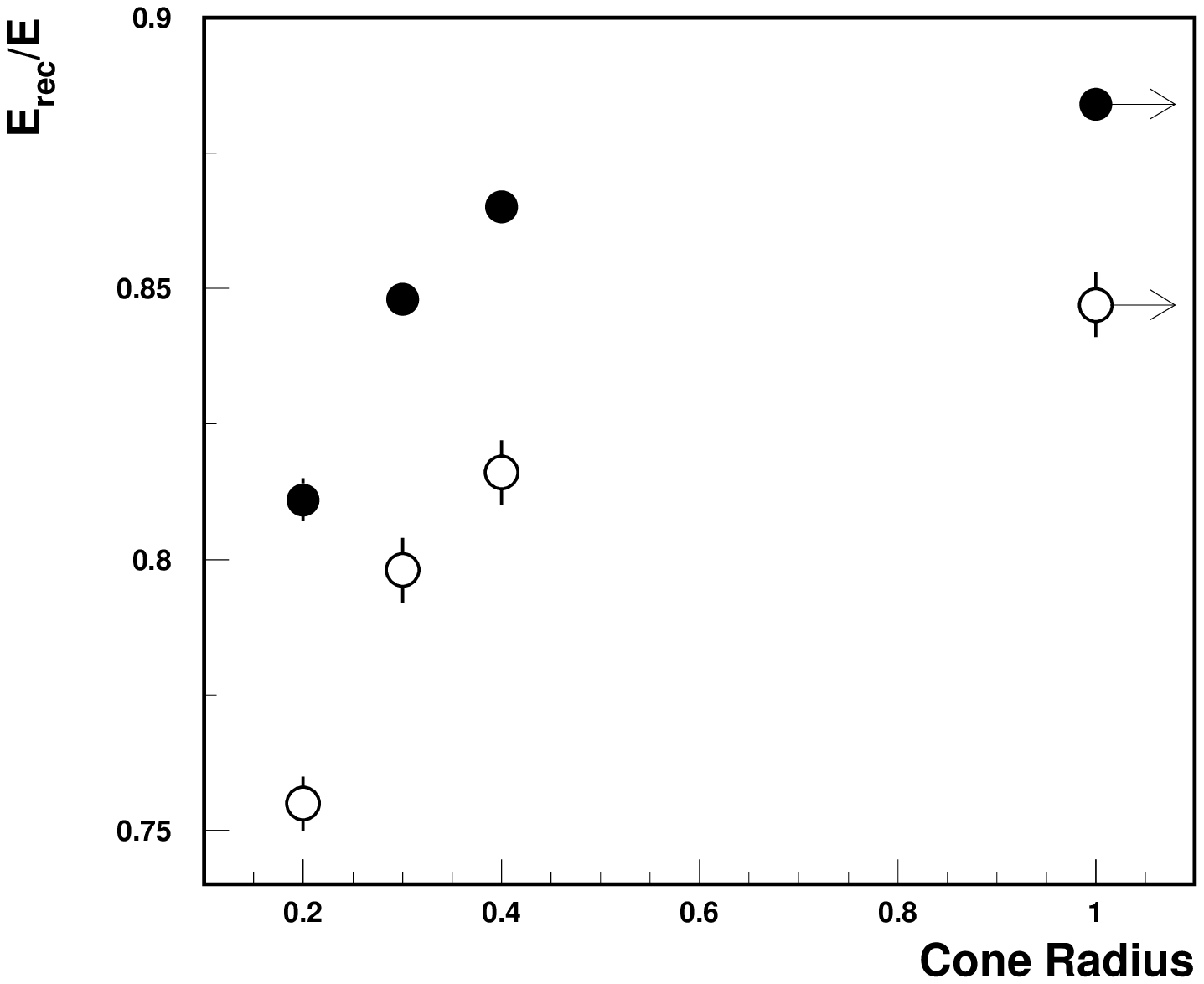,width=0.8\textwidth,height=0.4\textheight}}
\\
\end{tabular}
\end{center}
 \caption{
        Energy response for 50 GeV (open circles) and 200 GeV (solid dots)
        charged pions at $\eta = 0.3$ (top plot) and $\eta = 1.3$ (bottom plot)
        as a function of the cone size.
        The points with arrows correspond to the case without 
        restriction to a cone.
        Energy and pseudorapidity independent parameters were used for 
        the energy reconstruction.
        }
\label{f-06}
\end{figure}

\begin{figure}[tbph]
\begin{center}
\begin{tabular}{c}
\mbox{\epsfig{figure=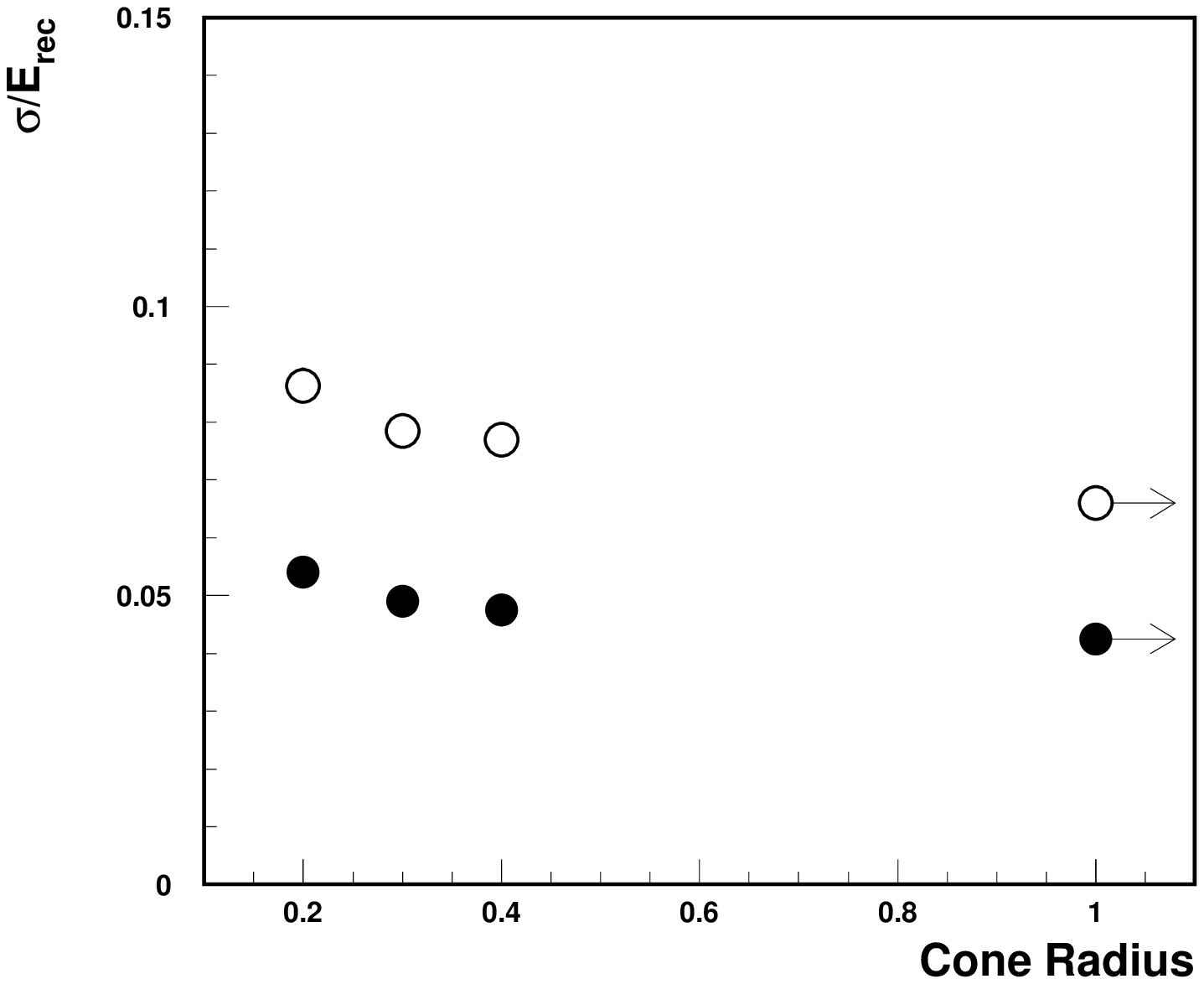,width=0.8\textwidth,height=0.4\textheight}}
\\
\mbox{\epsfig{figure=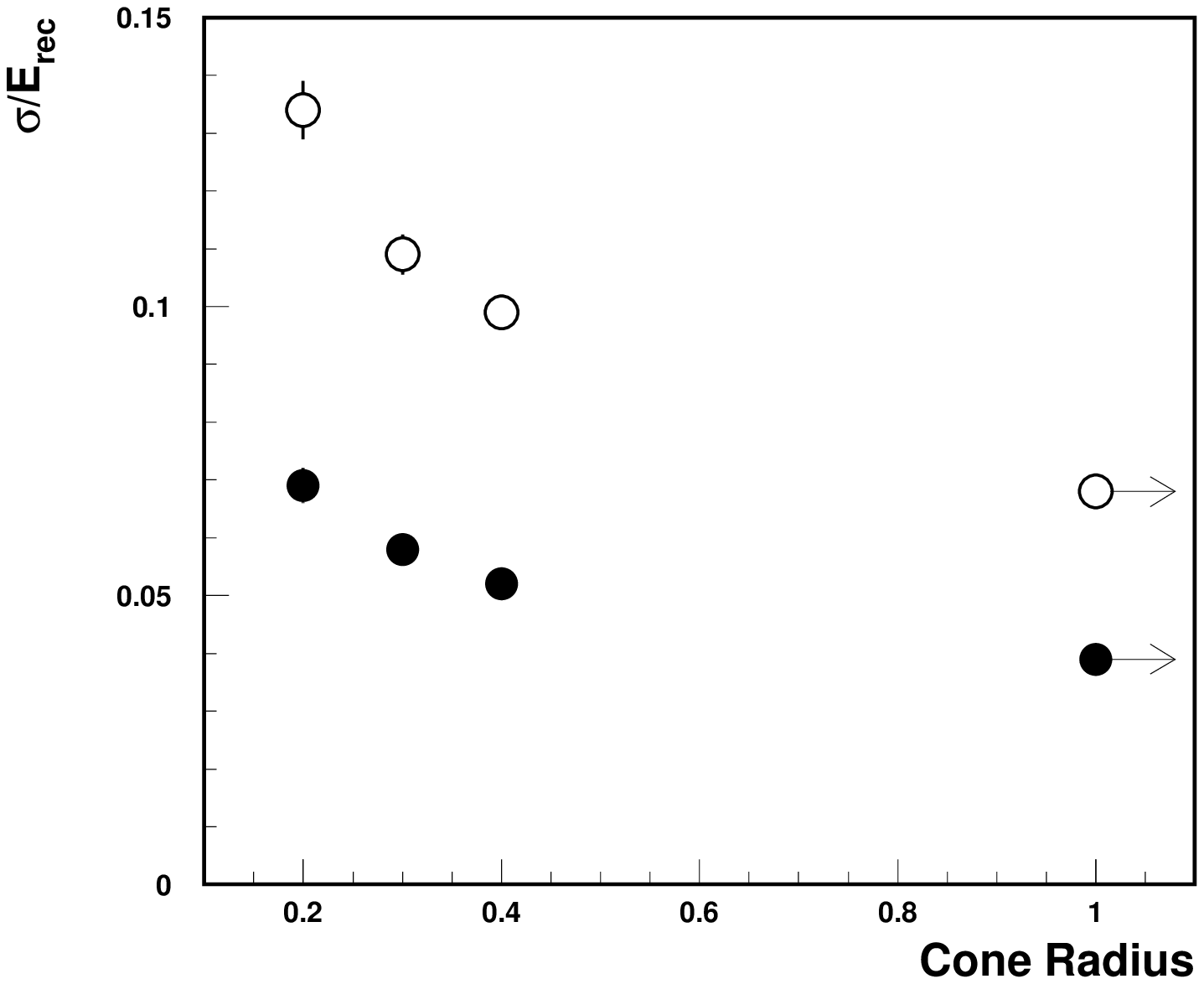,width=0.8\textwidth,height=0.4\textheight}}
\\
\end{tabular}
\end{center}
 \caption{
        Energy resolution for 50 GeV (open circles) and 200 GeV (solid dots)
        charged pions at $\eta = 0.3$ (top plot) and $\eta = 1.3$ (bottom plot)
        as a function of the cone size.
        The points with the arrows correspond to the case without 
        restriction to a cone.
        Energy and pseudorapidity dependent parameters were used for 
        the energy reconstruction.
        }
\label{f-07}
\end{figure}

The response and the energy resolution in the barrel region are presented 
in Figures \ref{f-06} and \ref{f-07} 
as a function of the cone size used for the pion energy reconstruction. 
Energy losses outside a cone noticeably increase 
with decreasing cone size, especially for 50 GeV pions. 
The energy resolution also becomes worse, but it is still 
acceptable for the cone of $\Delta R = 0.3$.   

\begin{figure}[tbph]
\begin{center}
\begin{tabular}{c}
\mbox{\epsfig{figure=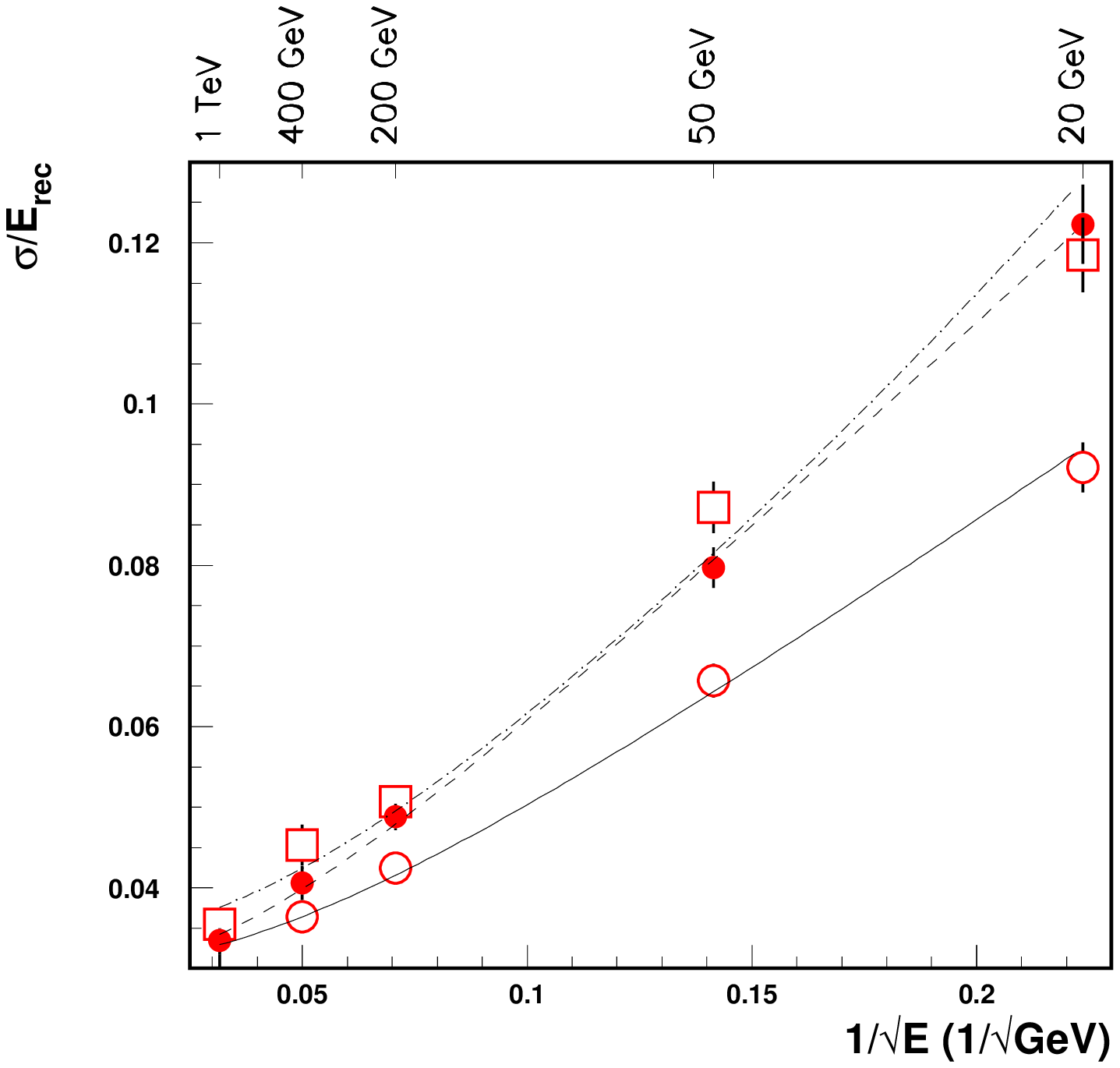,width=0.72\textwidth,height=0.36\textheight}}
\\
\mbox{\epsfig{figure=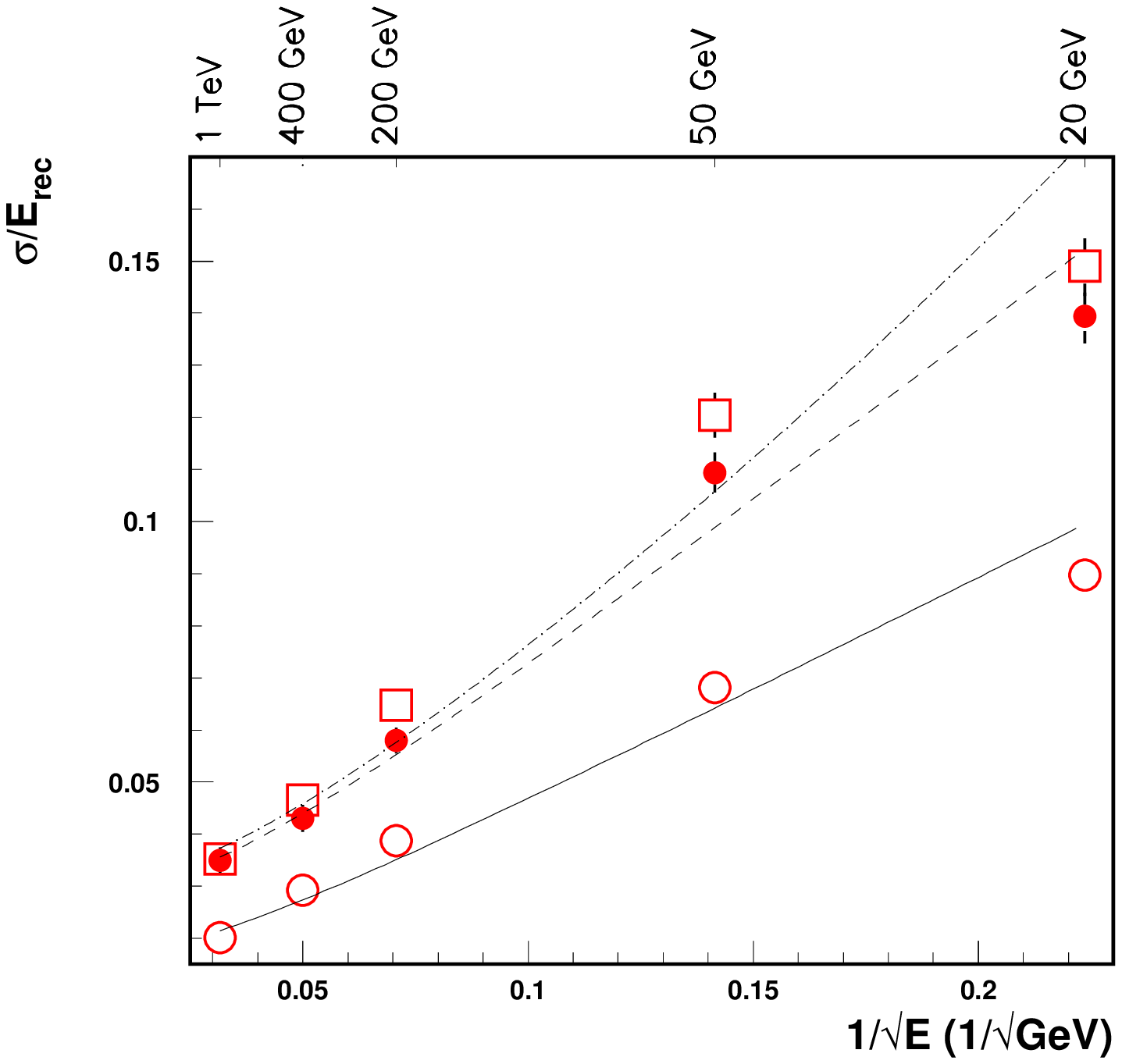,width=0.72\textwidth,height=0.36\textheight}}
\\
\end{tabular}
\end{center}
 \caption{
        Energy dependence of the resolution for pions at $\eta =0.3$
        (top plot) and $\eta = 1.3$ (bottom plot).
        The open circles show the intrinsic calorimeter resolution
        obtained with no cut cone restriction 
        and without electronic noise.
        The solid dots show results obtained for a cone of $\Delta R = 0.3$
        and without electronic noise.
        The open squares show the results obtained for a cone of 
        $\Delta R = 0.3$ and with a $2 \sigma$-noise cut (whit electronic
        noise included). The curves show the results of the fits 
        done with the
        two-term formula (\ref{eq-05}) 
        for the first two sets and with the tree-term formula (\ref{eq-06})
        for the third set.
        }
\label{f-08}
\end{figure}

Selecting cells with energy deposition above a certain threshold decreases 
the noise contribution. 
A study to optimise the cone 
size and the noise cut was performed in the barrel region.
A $2 \sigma$-noise applied to the calorimeter cells within a cone 
of $\Delta R = 0.3$ leads to the best energy resolution.
In Figure \ref{f-08}, 
the energy dependency of the resolution is plotted for two 
pseudorapidities: $\eta = 0.3$ and $\eta = 1.3$. 
The energy dependence of the resolution can be parametrised by the
equation \ref{eq-04} with an additional noise term:
\begin{equation}
\label{eq-06}
        \sigma /E = 
                a/\sqrt{E} \oplus b \oplus c/E \ 
\end{equation}
where $c$ is given in $\%GeV$. 
The results of the fit with the formula of Equation \ref{eq-06}
are presented in Table \ref{tb-02}. 
%2
\noindent
\begin{table}[tbph]
\begin{center}
\caption{
        Terms of the pion energy resolution 
        fitted with the two-term (\ref{eq-04})
        and the tree-term (\ref{eq-06}) expression
        for the $\eta = 0.3$ and $1.3$.
        }
\label{tb-02}
\begin{tabular}{|c|c|c|c|c|c|}
\multicolumn{6}{l}{\mbox{~~~}}          \\
\hline
 \multicolumn{3}{|c|}{Central  barrel region}
&\multicolumn{3}{|c|}{Extended barrel region}\\
 \multicolumn{3}{|c|}{$\eta = 0.3$}
&\multicolumn{3}{|c|}{$\eta = 1.3$}\\
\hline
$a$&$b$&$c$&$a$&$b$&$c$\\ 
($\% \ \sqrt{GeV}$)&($\%$)&($GeV$)&($\% \ \sqrt{GeV}$)&($\%$)&($GeV$)\\ 
\hline
\hline
\multicolumn{6}{|c|}{No Cone, No Noise}\\
\hline
$40\pm1$&$3.0\pm0.1$&--&$44\pm3$&$1.6\pm0.3$&--\\
\hline
\hline
\multicolumn{6}{|c|}{Cone $\Delta R = 0.3$, No Noise}\\
\hline
$53\pm2$&$3.0\pm0.2$&--&$67\pm4$&$2.9\pm0.4$&--\\
\hline
\hline
\multicolumn{6}{|c|}{Cone $\Delta R = 0.3$, Noise with a $2 \sigma$-cut}\\
\hline
$50\pm4$&$3.4\pm0.3$&fixed &$68\pm8$&$3.0\pm0.7$&fixed\\
        &           &at 1.0&        &           &at 1.5\\
\hline
\end{tabular}
\end{center}
\end{table}

\vspace*{-10mm}
%%%%%%%%%%%%%%%%%%%%%
\section{Conclusions}
%%%%%%%%%%%%%%%%%%%%%
The response of the barrel and extended barrel region of the 
ATLAS calorimeter system  to single 
charge pions was investigated using full simulation.
Pion energy scans from $E = 20$ GeV to $1000$ GeV  
 and pseudo rapidity scans  with pions of constant
transverse energy ($E_T = 20$ and $50$ GeV) have been analysed. 
 For the pion energy reconstruction, the ''Benchmark approach'' was used:
it provides a first order correction for non-compensation effects and
accounts for the effect of the dead material by 
using the ITC's and scintillators to sample the energy loss or
 interpolating between the energy deposited in adjacent calorimeter layers.

Energy and rapidity dependent and independent calibrations have 
been considered. The best results are obtained with energy and rapidity
 dependent parameters. The effect of electronic noise has been studied:
cone size and cell energy cuts have been optimised.
The energy dependence of the resolution can be parameterized as: 
$(50\pm4)\%/\sqrt{E} \oplus (3.4\pm0.3)\% \oplus 1.0/E$ at $\eta = 0.3$ and
$(68\pm8)\%/\sqrt{E} \oplus (3.0\pm0.7)\% \oplus 1.5/E$ at $\eta = 1.3$. 
The larger constant term at $\eta=0.3$ can be explained by the longitudinal 
leakage from calorimeters in this region. 
The resolution, obtained for the  pseudorapidity scans, is represented by: 
$(39\pm1)\%/\sqrt{E} \oplus (1\pm5)\%$ for $E_T = 20$ GeV,
$(49\pm9)\%/\sqrt{E} \oplus (2\pm6)\%$ for $E_T = 50$ GeV,
in the full range, except from about  $|\eta| = 1.3$ to 
$|\eta| = 1.5$, where the resolution is 
deteriorated by the energy loss in the dead material although  
no significant tails in the energy spectrum appears.

\vspace*{-5mm}
%%%%%%%%%%%%%%%%%%%%%%%%%%
\section*{Acknowledgements}
%%%%%%%%%%%%%%%%%%%%%%%%%%
The authors are grateful to Andrei Kiryunin for fruitful discussions. 

%%%%%%%%%%%%
%References
%%%%%%%%%%%%

\vspace*{-5mm}
%%%%%%%%%%%%%%%%%%%%%%%%%%
%  bibliography
%%%%%%%%%%%%%%%%%%%%%%%%%%

%%%%%%%%%%%%%%%%%%%%%

%%%%%%%%%
%Figures
%%%%%%%%%

%%%%%%%%%%%%%%%%%%%%%%%%%%%%%%%%%%%%%%%%%%%%%%%%%%%%%%%%%%%%%%%%%%%
\newpage

%%%%%%%%%%%%%%%%%%%%%%%%%%%%%%%%%%%%%%%%%%%%%%%%%%%%%%%%%%%%%%%%%%%
%0
\clearpage

%%%%%%%%%%%%%%%%%%%%%%%%%%%%%%%%%%%%%%%%%%%%%%%%%%%%%%%%%%%%%%%%%%%
%1
\clearpage
%%%%%%%%%%%%%%%%%%%%%%%%%%%%%%%%%%%%%%%%%%%%%%%%%%%%%%%%%%%%%%%%%%%
%2
\clearpage

%%%%%%%%%%%%%%%%%%%%%%%%%%%%%%%%%%%%%%%%%%%%%%%%%%%%%%%%%%%%%%%%%%%
%3
\clearpage

%%%%%%%%%%%%%%%%%%%%%%%%%%%%%%%%%%%%%%%%%%%%%%%%%%%%%%%%%%%%%%%%%%%
%4
\clearpage
%%%%%%%%%%%%%%%%%%%%%%%%%%%%%%%%%%%%%%%%%%%%%%%%%%%%%%%%%%%%%%%%%%%
%4b
\clearpage

%%%%%%%%%%%%%%%%%%%%%%%%%%%%%%%%%%%%%%%%%%%%%%%%%%%%%%%%%%%%%%%%%%%
%5
\clearpage

%%%%%%%%%%%%%%%%%%%%%%%%%%%%%%%%%%%%%%%%%%%%%%%%%%%%%%%%%%%%%%%%%%%
%6
\clearpage

%%%%%%%%%%%%%%%%%%%%%%%%%%%%%%%%%%%%%%%%%%%%%%%%%%%%%%%%%%%%%%%%%%%
%7
\clearpage

%%%%%%%%%%%%%%%%%%%%%%%%%%%%%%%%%%%%%%%%%%%%%%%%%%%%%%%%%%%%%%%%%%%
%8
\clearpage

%%%%%%%%%%%%%
\end{document}